\documentclass[preprint,a4paper]{raa}
\usepackage{mathptmx}
\usepackage[utf8]{inputenc}
\usepackage{graphicx} 
\usepackage{natbib}
\usepackage{adjustbox} 

\usepackage{booktabs}   
\usepackage{siunitx}    
\usepackage{tabularx}   
\usepackage{physics}    
\usepackage{upgreek}    
\usepackage{multirow}
\usepackage{soul}
\usepackage{xcolor}  
\usepackage{amssymb,amsmath}

\newcommand{\gag}{g_{a\gamma\gamma}}
\bibpunct{(}{)}{;}{a}{}{,}
\usepackage[pagebackref=true]{hyperref}

\usepackage{xcolor}

\begin{document}
 
\title{Searching for axion dark matter conversion spectral lines in neutron star magnetospheres with FAST}
 
   \volnopage{Vol.0 (20xx) No.0, 000--000}      
   \setcounter{page}{1}         
    
    \author{Sinuo Gao\inst{1,2} \and Chen Wang\inst{2,3,4}\thanks{Corresponding Author: wangchen@nao.cas.cn} \and Maoyuan Liu\inst{1}}

    \institute{Department of Physics, College of Science, Tibet University, Lhasa 850000, China
    \and 
    National Astronomical Observatories, Chinese Academy of Sciences, Jia-20 Datun Road, ChaoYang District, Beijing 100012, China;{\it wangchen@nao.cas.cn} 
    \and
    School of Astronomy and Space Science, University of Chinese Academy of Sciences, Beijing 100049, China 
    \and
    Key Laboratory of Radio Astronomy and Technology,  Chinese Academy of Sciences, Beijing 100101, China\\ 
\vs\no
   {\small Received 20xx month day; accepted 20xx month day}
   }

\abstract{The axion is a well-motivated dark matter candidate, which can convert into narrow radio spectral lines via the Primakoff effect in the strongly magnetized magnetospheres of neutron stars. This provides a novel astrophysical probe for axion searches that is complementary to laboratory experiments. Using FAST, the world’s most sensitive single-dish radio telescope, we observed two X-ray dim isolated neutron stars (RXJ1605.3+3249 and RXJ1308.6+2127) within its sky coverage, which are predicted to yield the strongest axion-conversion spectral lines (ACL). Although no significant signal was detected at the $5\sigma$ confidence level, we establish a new upper limits on the axion-photon coupling constant 
$\gag \lesssim 5\times 10^{-12} \text{GeV}^{-1}$for axion masses ranging from 4.14 to 6.20\,$\mu$eV corresponding to the 1.0--1.5GHz observational band. This result constitutes the tightest constraint in this axion mass range among all existing studies employing the same ACL-based method.
\keywords{(cosmology:) dark matter - methods: observational - stars: neutron - radio lines: general}
}

\authorrunning{S. Gao, C. Wang \& M. Liu }            

\titlerunning{Search for axion conversion lines with FAST}

\maketitle

\section{Introduction}
\label{sect:intro}
The axion was originally proposed to resolve the strong CP problem in quantum chromodynamics (QCD) \citep{PQ1977}. As a light, long-lived pseudoscalar particle, the axion also represents a well-motivated cold dark matter candidate that contributes to the dominant mass component of the observable Universe, and has thus become a major focus of modern particle cosmology. The coupling between axion dark matter and photons is described by the interaction Lagrangian $\mathcal{L} = \gag a \mathbf{E} \cdot \mathbf{B}$, where $\mathbf{E}$ and $\mathbf{B}$ denote the electric and magnetic fields, respectively,  $a$ represents the axion dark matter field, and $\gag$ is the axion-photon coupling parameter \citep{IR2018}. For the QCD axion, $\gag$ is directly related to the axion mass, whereas this relation is relaxed for axion-like particles (ALP). This distinctive property provides a unique observational probe for axion searches. 

Numerous ground-based experiments have been carried out to explore the axion-photon coupling parameter space. These include experiments using microwave cavity resonators to search for axion-to-photon conversion in the galactic dark matter halo, such as RBF \citep{WdS1989}, ADMX \citep{BBB2021}, and CAPP \citep{AAA2024}, et al. Another class of experiments employs the helioscope technique to hunt for solar axions, with representative examples including CAST \citep{CAA2017} and IAXO \citep{CSX2020}. Additional laboratory-based searches include ALPS \citep{DGG2022}, which employs the light-shining-through-a-wall method, and PVLAS \citep{DEG2016}, which searches for axion-induced vacuum polarization and birefringence in strong magnetic fields.
To date, these experiments have not detected conclusive evidence for axions, but have placed stringent upper limits on the axion parameter space 
\citep{FB2008,EFG2010,BDF2015}.

Complementary to laboratory searches, various astrophysical and cosmological observations can also be used to constrain axion or ALP parameters. Examples include the $\gamma$-ray transparency of distant astrophysical sources \citep{SHS2008,AGR2011}, the soft X-ray excess in galaxy clusters \citep{CM2013}, and the horizontal-branch to red-giant star ratio in Galactic globular clusters. All of these probes rely on potential axion-photon conversion effects to constrain the coupling parameter.

In recent years, a highly promising observational pathway has emerged: dark matter axions can convert into photons via the Primakoff effect in the ultra-strong magnetospheres of neutron stars \citep{PP2009}. Resonant axion-photon conversion occurs when the axion mass matches the local plasma frequency ($m_a = h\nu_p$) inside the pulsar magnetosphere, leading to efficient conversion into radio photons. The conversion efficiency scales positively with the magnetic field strength at the resonance point \citep{PP2009,HKS2018,Huang2018}. The resulting radio signal appears as a narrow spectral feature known as axion conversion spectral line (ACL), whose central frequency is determined by the axion mass. The radiated power of ACL is given by \citep{WNE2021,BGM2021,ZHJ2022} 
\begin{equation} \label{eq:ACLpower}
\begin{aligned}
\frac{{\rm d}P}{{\rm d}\Omega}\approx & \,5.7\times10^9 \, \text{W}  
\left( \frac{\gag}{10^{-12} \, \text{GeV}^{-1}} \right)^2
\left( \frac{B_*}{10^{14} \, \text{G}} \right)^{5/6} 
\left( \frac{P}{1 \, \text{s}} \right)^{7/6}  \\ 
\times& 
\left( \frac{m_a}{1 \, \text{GHz}} \right)^{4/3} 
\left( \frac{\rho_\infty}{0.45 \, \text{GeV cm}^{-3}} \right) 
\left( \frac{v_0}{200 \, \text{km s}^{-1}} \right)^{-1} \\
\times& 
\left( \frac{R_{\text{NS}}}{10 \, \text{km}} \right)^{5/2} 
\left( \frac{M_{\text{NS}}}{1 \, M_\odot} \right)^{1/2} 
\frac{3(\hat{\mathbf{m}}\cdot\hat{\mathbf{r}})^2+1}{\left|3\cos{\theta}\,\hat{\mathbf{m}}\cdot\hat{\mathbf{r}}-\cos\theta_m\right|^{7/6}},
\end{aligned}
\end{equation}
where ${B_*}$ is the NS surface magnetic field strength, $P$ is the NS spin period, $R_{\text{NS}}\approx 10\text{km}$  is the neutron star radius, $M_\text{NS}\approx 1 \, M_\odot $  is the neutron star mass, $\rho_\infty\approx 0.45 \, \text{GeV cm}^{-3}$ is the local axion dark matter density far from the star, and $v_0\approx200 \, \text{km s}^{-1}$ is the Galactic dark matter velocity dispersion \citep{BT2012,Read2014}. Here $\theta_m$ and $\theta$ are the angles between NS spin axis and respectively, the magnetic dipole direction $\hat{\mathbf{m}}$  and  the line of sight  $\hat{\mathbf{r}}$. 

The ACL width is dominated by Doppler broadening from propagation in the magnetospheric plasma, with a relative width $\delta \nu / \nu \sim 5 \times 10^{-6}$  \citep{FKM2020}. Assuming a Gaussian profile with intrinsic width $\sigma_0=5\times10^{-6}m_a$,  the peak flux density of ACL is 
\begin{equation} \label{eq:ACL}
\begin{aligned}
S_{\text{ACL}} &= \frac{{\rm d}P/{\rm d}\Omega}{d^2}\frac{1}{\sqrt{2\pi}\sigma_0}\approx \,4.8\times10^{-6} \, \text{Jy} \\
  &\times \left(\frac{{\rm d}P/{\rm d}\Omega}{5.7\times10^9 \, \text{W}}\right) 
  \left( \frac{d}{100 \, \text{pc}} \right)^{-2} 
  \left( \frac{m_a}{1 \, \text{GHz}} \right)^{-1},
\end{aligned}
\end{equation}
with $d$ the distance to the neutron star. The average flux density in frequency channel $i$ is then given by \citep{ZHJ2022}
\begin{equation} \label{eq:Snu}
\bar{S}_{\nu_i}(m_a,\gag)=\frac{S_{\rm ACL}}{\Delta\nu}\int_{\nu_{i,\rm min}}^{\nu_{i,\rm max}}e^{-\frac{(\nu-m_a)^2}{2{\sigma_0}^2}} {\rm d}\nu.
\end{equation}
where $\Delta\nu$ is the channel width, $\nu_{i,\rm min}$ and $\nu_{i,\rm max}$ denotes the lower and upper frequencies of the $i$ channel.  

Recent theoretical developments have included more realistic treatments of the neutron star magnetospheric structure, emission geometry, plasma density gradients, and general relativistic effects \citep{WNE2021,BGM2021,XLG2023}. By implementing ray-tracing simulations and modeling periodic flux variations and polarization signatures, the detectability of ACLs can be significantly improved, allowing tighter constraints on $\gag$, albeit with an increased dependence on magnetospheric models.

X-ray Dim Isolated Neutron Stars (XDINSs), a nearby class of neutron stars, possess the combined characteristics of ultra-strong magnetic fields ($>10^{13}$ G), long spin periods (3--17 s), and proximity (160--500 pc) \citep{H2007,K2008}. 
They exhibit purely thermal X-ray emission with no detectable radio pulsations, which greatly reduces contamination to potential axion-induced radio lines, making them prime targets for axion searches. In addition, some nearby magnetars and normal pulsars with extreme magnetic fields may also produce detectable ACL emission. The Galactic Center magnetar J1745$-$2900 is especially promising, as the dark matter density in its vicinity is predicted to be extremely high \citep{NFW1997}, which can strongly enhance the expected axion flux. In recent years, multiple observational searches have been performed for theoretically predicted ACLs from neutron stars (including XDINSs and pulsars) using high-sensitivity radio telescopes and interferometric arrays \citep{D2020A,D2020P,FKM2020,ZHJ2022,BKM2023,LCG2025}.

In this work, we present observations of two XDINSs using the Five-hundred-meter Aperture Spherical Telescope (FAST), the most sensitive single-dish radio telescope currently operating. We have obtained the most stringent constraints on the coupling constant archieved by the ACL-based method to date, in the axion mass range covered by our FAST bandwidth. This paper is organized as follows. Section 2 describes the observations and data processing procedures. Section 3 presents our main results and the constraints on the axion parameter space. Section 4 summarizes our findings and provides a discussion.

\section{Observations and Data Reduction}
\label{sec2}
\subsection{Observation Design}

The Five-hundred-meter Aperture Spherical radio Telescope \citep{NAN2008,NLJ2011,JTH2020} has the largest collecting area for radio waves, with an aperture of 300 m in diameter. Mounted together with the 19-beam L-band receiver that has a system temperature of about 20 K and a gain $G\approx 16$ K/Jy, FAST is currently the most sensitive radio telescope to search the possible ACL in NS. 

Using Eq.~\ref{eq:ACLpower} and \ref{eq:ACL}, we could calculate the theoretical peak flux density of ACL by assuming $\gag = 10^{12} \, \text{GeV}^{-1}$, $m_a=1$\,GHz, $\theta=\pi/2$ and $\theta_m=0$ for all candidate NSs (including XDINSs, magnetars, and pulsars) with known $P$, $B_*$ and $d$, to select the targets with the highest ACL flux density. The targets should be within the FAST-visible sky area ($-13^\circ < \text{DEC} < 66^\circ$). The two XDINSs, RXJ1605.3+3249 and RXJ1308.6+2127,  with highest $S_\text{ACL}$ (see Table~\ref{tab:2XDINSs}) were chosen as prime targets for FAST observational campaigns to detect ACL signatures or constrain coupling constant.

\begin{table}[htbp]
  \centering
  \caption{The two XDINSs with highest predicted $S_\text{ACL}$ in the FAST-visible sky .\label{tab:2XDINSs}}

  \begin{tabular}{cccccc}
    \toprule
    {Source} & {$P$} & {$d$} & {$B_*$} & {$S_\text{ACL}$} & {Ref.$^a$} \\
    & {(\si{s})} & {(\si{kpc})} & {(\si{G})} & {(\si{\micro  Jy})} &\\
    \midrule
    \textbf{RXJ1605.3+3249} & 6.88$^b$ & 0.390 & 9.00E13 & 2.74 & {1, 2}\\
    \textbf{RXJ1308.6+2127} & 10.30 & 0.525 & 3.40E13 & 1.07 & {1, 3, 4}\\
    \bottomrule
  \end{tabular}
  
  \vspace{0.2cm}
  \raggedright
  \footnotesize{Notes:\\  
  $^a$ Reference: 1. \citealt{H2007}; 2. \citealt{MBH2019}; 3. \citealt{BH2024}; 4. \citealt{PPH2007}\\
  $^b$ We note that the latest work of \citealt{MBH2019} did not detect a spin period for this source at sufficient significance, so we still adopt the period from \citealt{H2007} here.}
\end{table}

The observational frequency band of the 19-beam L-band receiver on the Five-hundred-meter Aperture Spherical Telescope (FAST) is 1.0--1.5 GHz. Its spectral backend is configured with 1024k frequency channels and a frequency resolution of 0.48 kHz, and both linear polarizations were recorded during observations.

Observations of the two X-ray Dim Isolated Neutron Stars (XDINSs) adopted the ``On-Off'' position switching mode: a 5-minute ``On'' phase, during which the M01 beam tracked the target source, followed by a 5-minute ``Off'' phase, during which the same beam tracked an adjacent cold sky region offset from the source. This cycle was repeated continuously to establish a reliable reference baseline for subsequent data analysis, enabling accurate subtraction of background emission through On-Off comparison.
Signals from the remaining 18 beams (M02--M19) were also recorded simultaneously, serving as auxiliary data for subsequent cross-comparison and radio frequency interference (RFI) excision.

Observations of RXJ1605.3+3249 were split into two segments: 54.5 minutes on 20 July 2021 and 197.5 minutes on 3 January 2021, resulting in a total of 4.2 hours of valid data. Observations of RXJ1308.6+2127 were performed on 17 December 2020, with a total of 2.2 hours of valid data.

\subsection{Data reduction}
\label{sect:data} 
\begin{figure}
    \centering
    \includegraphics[width=1\linewidth]{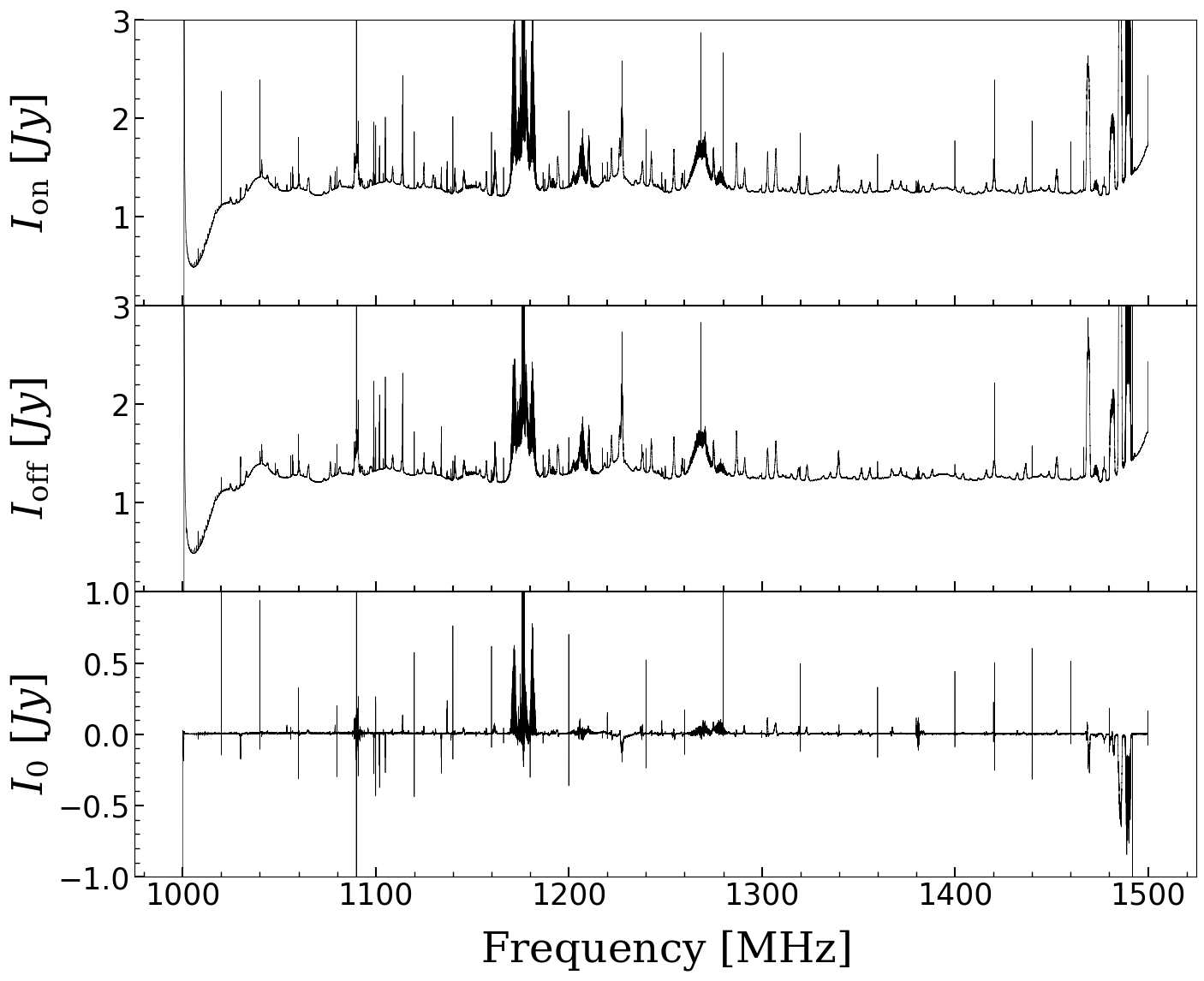}
    \caption{The calibrated spectrum of RXJ1605.3+3249, including the average spectrum of On-source phases $I_\text{on}$ (top panel), Off-source phases $I_\text{off}$ (middle), and their difference $I_0=I_\text{on}-I_\text{off}$.}
    \label{fig:cal}
\end{figure}

During observations, periodic noise signals with a duration of 20 seconds were injected every 5 minutes via a noise diode for data calibration. The modulated noise signal had a period of 2 seconds (1 second ON and 1 second OFF), with an approximate amplitude of 1 K; its frequency dependence was provided by the FAST operation team. For each frequency channel, the noise response value was derived by averaging the difference between the ON and OFF states of the noise signal. Combined with FAST gain measurements from published literature \citep{grs+22,chl+24,lwj+24} and data provided by the FAST operation team, a frequency-dependent calibration coefficient was obtained for each observation run. This coefficient converted the raw response value of each channel into the calibrated flux density in Jansky (Jy). The two linearly polarized flux signals were calibrated individually and then summed to yield the total flux spectrum. Figure \ref{fig:cal} presents the calibrated spectra of the On-source and Off-source phases for RXJ1605.3+3249(observed on 3 January 2021), as well as their difference $I_0=I_\text{on}-I_\text{off}$ with most of the background interference and instrumental drifts subtracted.

Despite the aforementioned processing, residual systematic artifacts—primarily baseline non-flatness—still obscure potential axion signals. We derived the baseline \ $\mu$ by smoothing the spectrum with a Savitzky-Golay filter, using a smoothing window of approximately 60 kHz. This window width is significantly wider than the typical ACL width of 5.0--7.5 kHz within the 1.0--1.5 GHz observing band, ensuring that the ACL signal is not smoothed out during baseline estimation.

\begin{figure}
    \centering
    \includegraphics[width=1\linewidth]{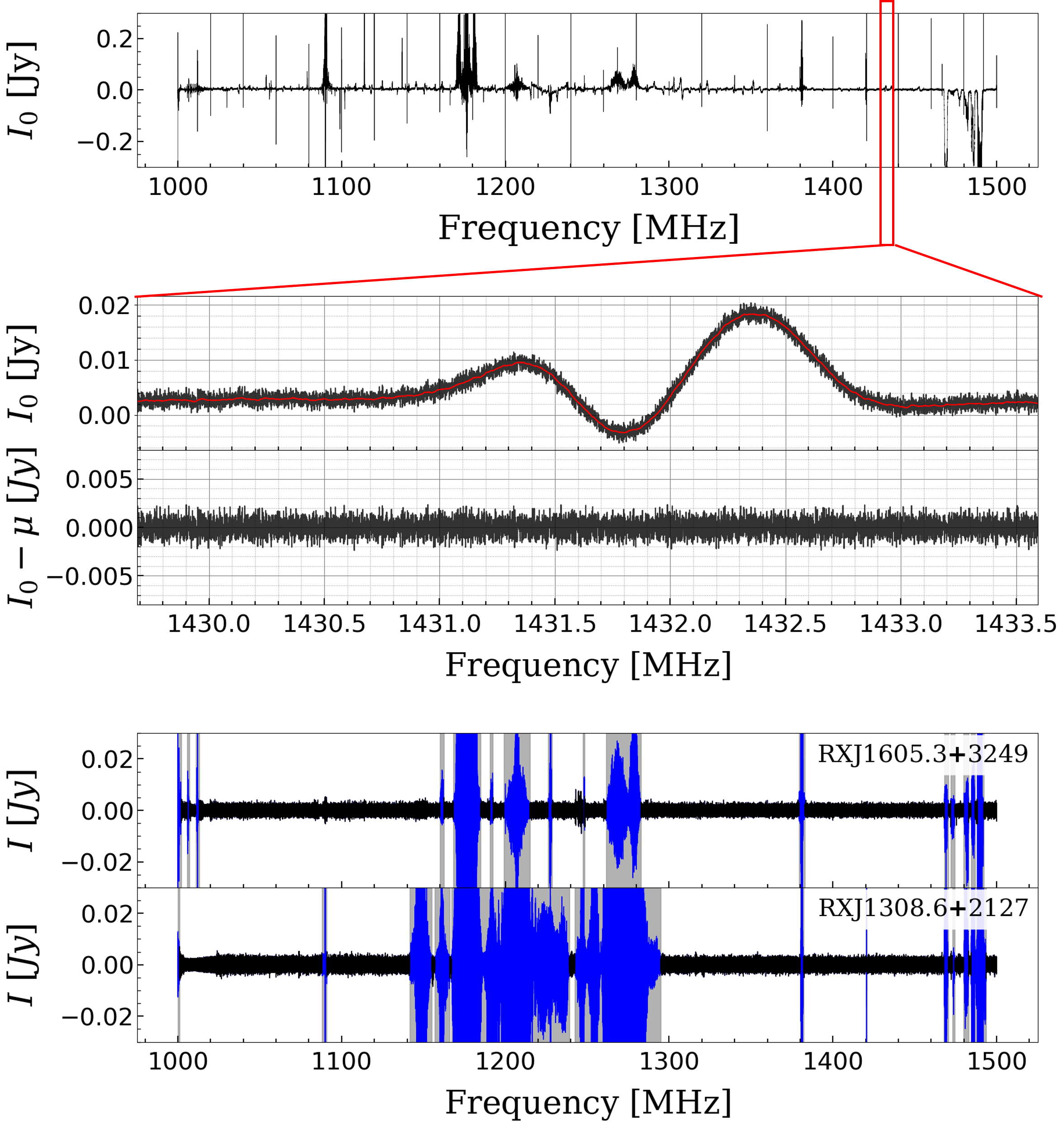}
    \caption{Baseline subtraction and RFI excision procedure.
    Top panel: original spectrum $I_0$ of RXJ1605.3+3249.
    Middle panel: baseline $\mu$ (red curve) fitted with a Savitzky-Golay filter, and the baseline-subtracted spectrum $I_0-\mu$. A small segment of the spectrum is shown for illustration.
    Bottom panel: clean spectrum $I(\nu)$ after baseline subtraction and RFI excision for the two XDINSs. Excised RFI bands are indicated by gray regions with blue curves.}
    \label{fig:baseline_remove}
\end{figure}

The spectrum after baseline subtraction ($I_0-\mu$) still contained substantial radio frequency interference (RFI), which would severely contaminate subsequent axion searches and thus required excision. Fortunately, signals from all 19 beams of the FAST L-band receiver were recorded simultaneously during observations. In general, RFI manifests in the same frequency bands across the auxiliary beams (M02--M19), which greatly facilitates RFI identification. For spurious interference present in only one or two frequency channels, the affected channels were replaced with random values having the same variance as the adjacent clean data. For continuous broadband interference, the corresponding affected frequency segment was directly excised. The total bandwidth of excised RFI amounts to approximately 85.0\,MHz for RXJ1605.3+3249 and 169.6\,MHz for RXJ1308.6+2127. The clean spectrum  $I(\nu)$ of two XDINSs after baseline subtraction and RFI excision is shown in the bottom panel of Figure \ref{fig:baseline_remove}, which was used for subsequent ACL line detection and constraints on the axion-photon coupling parameters.

\section{Data Analysis And Results}
\label{sect:analysis}
\subsection{ACL signal search}

Following calibration, baseline subtraction, and RFI excision, the matched filtering method was employed to extract potential axion conversion line (ACL) signals from the cleaned spectrum $I(\nu)$. The core principle of this method is to enhance signal energy and suppress background noise by convolving the cleaned spectrum with a predefined ACL template. For the targeted axion signals in this work, we adopted a Gaussian-shaped ACL filter template, given by
\begin{equation} \label{eq:ACL_template}
s_\nu = \frac{1}{\sqrt{2\pi}w}e^{-\frac{(\nu-m_a)^2}{2{w}^2}}.
\end{equation}
where $w$ denotes the template width. To cover the theoretically predicted ACL width range of 5.0--7.5 kHz, the template width $w$ was continuously incremented from 2.0 kHz to 10 kHz with a step size of 0.5 kHz, which matches the frequency resolution of FAST.
The standard deviation of the filtered response was defined as 
 $\sigma_{\text{filter}} = \sigma_I  \sqrt{\sum_{i=1}^{n} s_{i}^2}$
, where $\sigma_I$ is the standard deviation of the cleaned spectrum $I(\nu)$,  $s_{i}$ represents the value of the $i^{\text{th}}$ point in the matched filter template (Eq.~\ref{eq:ACL_template}), and $n$ is the total number of data points in the template.

The signal-to-noise ratio (SNR) of the matched filtering results was quantified in units of $\sigma_\text{filter}$. A threshold of $5\sigma$ was adopted to identify statistically significant signals. Matched-filter searches were performed on the processed data of the two XDINSs, RXJ1605.3+3249 and RXJ1308.6+2127. No signals exceeding the $5\sigma$ threshold were detected in either target.

\subsection{Constraints on the coupling constant}
Since no credible ACL lines were identified in the spectrum of the two XDINs, a Bayesian method was applied to set an upper limit on the coupling constant $\gag$. We first downbinned the high-frequency-resolution spectrum $I(\nu)$ from 0.48\,kHz to 4.8\,kHz  (scrunch 10 frequency channels to one channel), which concentrated most of the energy of ACL into a single frequency channel and significantly increased the computation speed. The likelihood function is given by:
\begin{equation} \label{eq:likelihood}
L(m_a,\gag)= \prod_{i=1}^{N} \frac{1}{\sqrt{2\pi}\sigma_{\nu_i}} \,\exp\left[-\frac{(I_{\nu_i}-\bar{S}_{\nu_i}(\boldsymbol{m_a,\gag})^2}{2\sigma_{\nu_i}^2}\right].  
\end{equation}
where $\nu_i$ denotes the index for the $i-$th frequency channel, $I_{\nu_i}=I(\nu_i)$ is the spectrum flux density of the $\nu_i$ channel, $\sigma_{\nu_i}^2$ is the variance of $I_{\nu_i}$. The total number of channels is $N=21$, corresponding to a frequency range of 100 kHz, which includes the central channel (equals $m_a$) and all the sideband channels with putative excess of $\bar{S}_{\nu_i}$. 

\begin{figure}
    \centering    
    \includegraphics[width=0.99\linewidth]{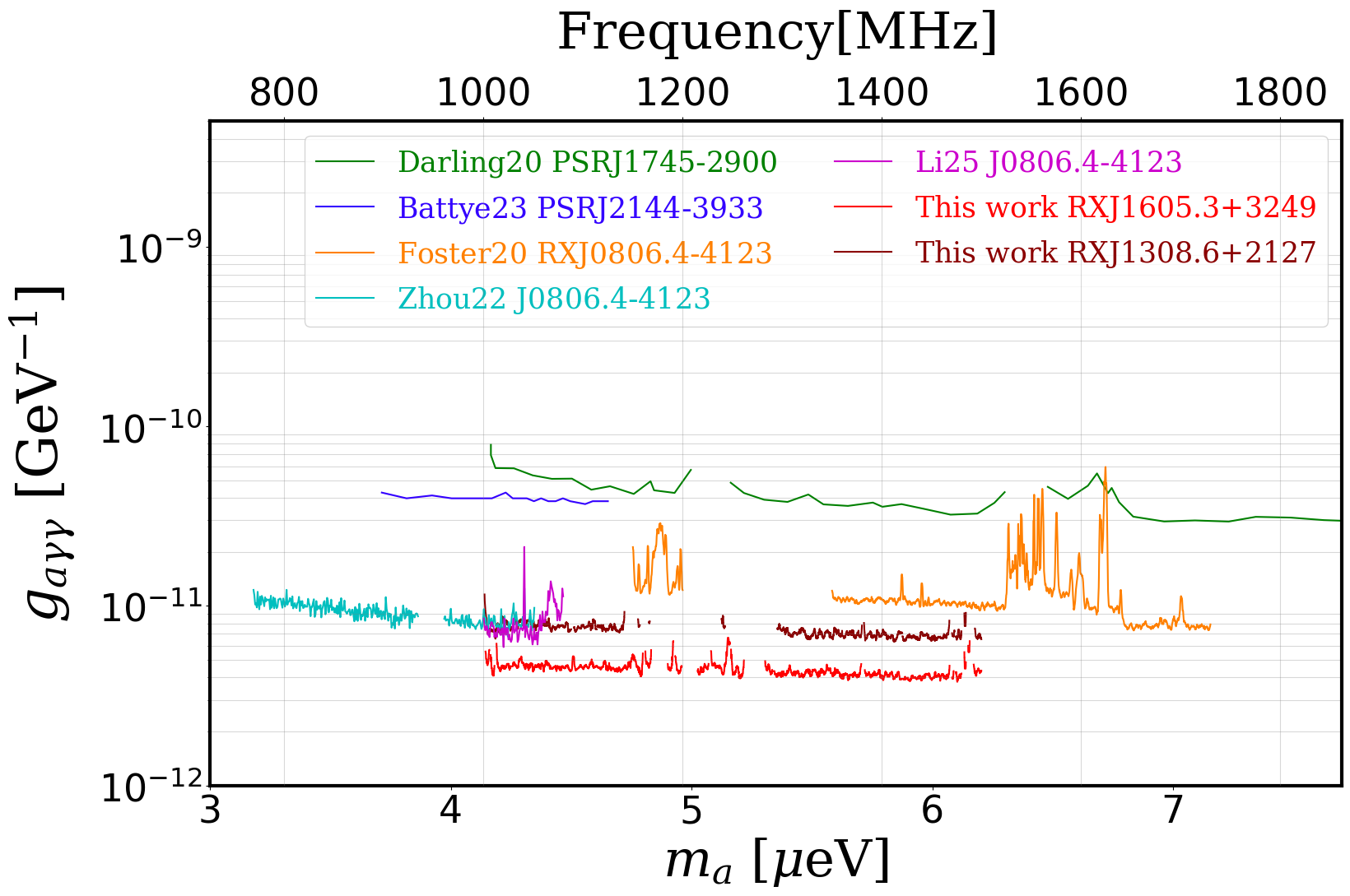}
    \includegraphics[width=1.0\linewidth]{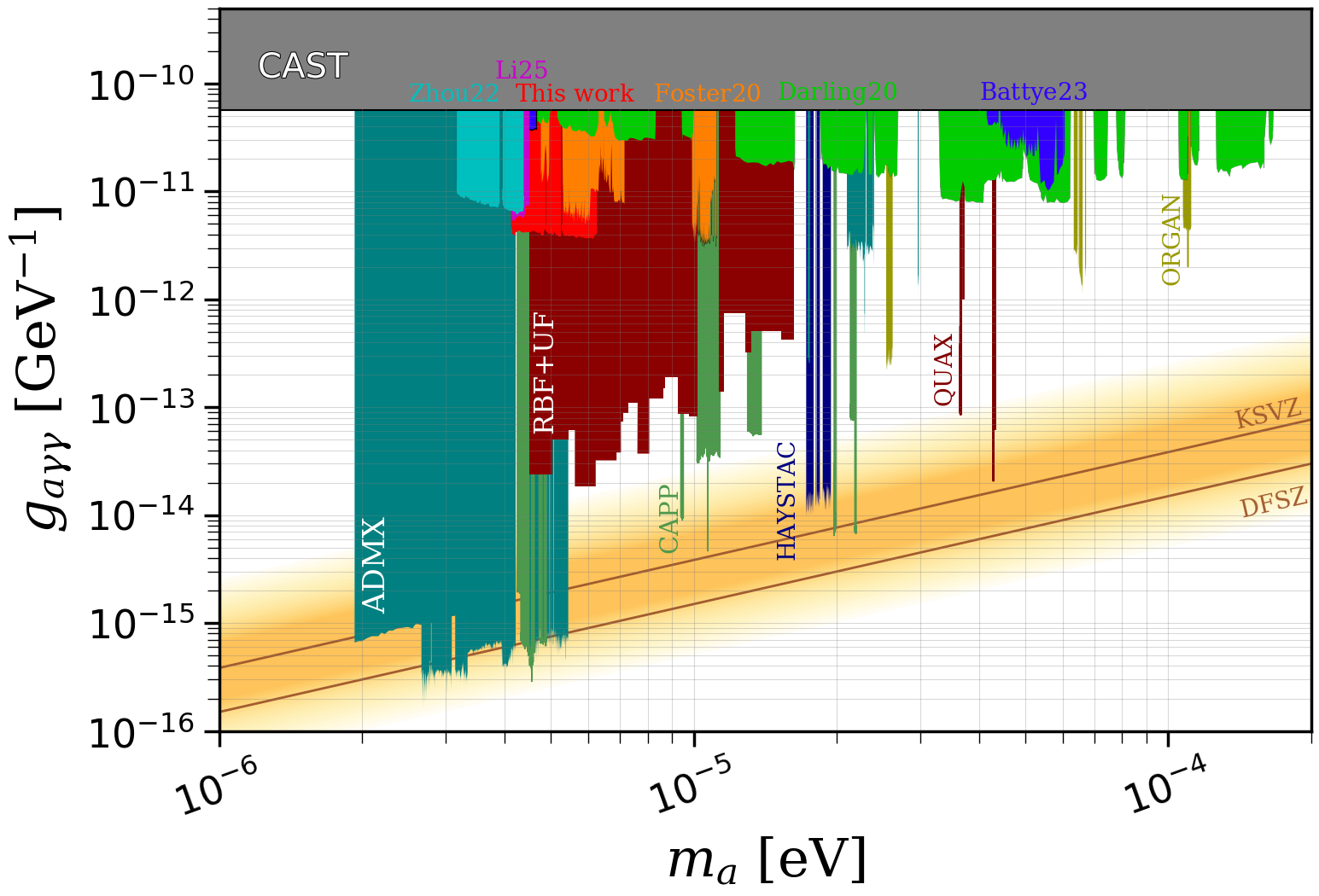}
    \caption{The 95$\%$ confidence level upper limits on $\gag$ as a function of $m_a$ derived from FAST observations of RXJ1605.3+3249 (red curve) and RXJ1308.6+2127 (dark-red curve), compared with other results.   The top panel focuses on the radio band region of 0.8-1.8\,GHz, including previous upper limits on $\gag$ from the ACL-based method \citep{D2020A,D2020P,BKM2023,FKM2020,ZHJ2022, LCG2025}. The bottom panel also shows results from laboratory experiments such as ADMX, RBF, CAPP, et al.\citep{BBB2021,APS1987,CAA2017,BCC2020,BBC2023,HSS1990,KLC2021,BPK2021,AAB2021,ABC2021,BGE2017}, obtained using the AxionLimits Code \citep{cohare2020}. }
    \label{fig:garr}
\end{figure}

We uniformly sampled $\mathrm{log}_{10}(\gag)$ in the range $[-13, -10]$.
The geometric parameters $\theta$ and $\theta_m$  are treated as nuisance parameters and marginalized over, assuming uniform priors in $[0, \pi]$ and $[0,\pi/2]$, respectively. For a given axion mass $m_a$, we evaluate the likelihood  $P(D_a|\gag,\theta,\theta_m)$ over the three-dimensional parameter space, then marginalize over  $\theta$ and $\theta_m$ to obtain $P(D_a|\gag)$. The posterior distribution of $\gag$ is obtained via Bayes’ theorem: $P(\gag|D_a)\propto P(D_a|\gag)P(\gag)$, where $P(\gag)$ is the prior distribution on the coupling constant.  

We then derived the  $95\%$ confidence level upper limit on $\gag$ from this posterior distribution.  We also examined the impact of different downbin numbers on the final constraint and found almost no variation. Our constraints, together with other observational and experimental results, are displayed in Figure~\ref{fig:garr}.  In the axion mass range of $m_a=4.14$--6.20\,$\mu$eV, corresponding to the FAST observation band 1.0--1.5\,GHz, we achieved a more stringent constraint on  $\gag$  than previous studies based on the ACL method.

\section{Conclusion and Discussion}
\label{sect:discussion}
Using observations of two XDINS (RXJ1605.3+3249 and RXJ1308.6+2127) with FAST, we searched for radio spectral lines converted from dark matter axions in the magnetospheres of neutron stars. No significant axion conversion line signal was identified at the $5\sigma$ confidence level. 
We derived an upper limit on the axion-photon coupling constant of  $\gag\lesssim 5\times10^{-12}\text{GeV}^{-1}$  for axion masses in the range  4.14--6.20\,$\mu$eV, representing the tightest constraint in this mass range among all existing results from the ACL‐based method.  Although contemporary laboratory experiments provided more stringent limits on the coupling constant (see the lower panel of Figure~\ref{fig:garr}), constraints from independent astrophysical methods remained a highly important and complementary detection strategy for axion searches.

Future longer observations could further improve the constraints on the axion-photon coupling constant. In addition, adopting more realistic magnetospheric plasma models for NSs, accounting for variations in line intensity and polarization across different NS rotation phases, and including general relativistic effects and radio wave refraction could lead to even tighter constraints on the coupling constant, although these treatments would also introduce additional model uncertainties. We left these investigations for future work.

\section*{Acknowledgements}
This work made use of the data from FAST (Five-hundred-meter Aperture Spherical radio Telescope, https://cstr.cn/31116.02.FAST). FAST is a Chinese national mega-science facility, operated by National Astronomical Observatories, Chinese Academy of Sciences. This work is supported by the National Natural Science Foundation of China (No. 12133004) and the National Key R\&D Program of China (No. 2025YFA1614003).  We would like to thank Hao Chen and Mengtian Li for kind support during this work.

\bibliographystyle{raa}  
\bibliography{bibtex}

\label{lastpage}

\end{document}